\documentclass[conference, letterpaper]{IEEEtran}

\ifCLASSINFOpdf
\else
\fi

\ifCLASSINFOpdf
   \usepackage[pdftex]{graphicx}
\else
\fi

\usepackage[cmex10]{amsmath}
\usepackage{color}
\usepackage{fancyhdr}
\usepackage{amssymb}
\usepackage{pifont}
\usepackage{graphicx}
\usepackage{epstopdf}
\usepackage{epsfig}
\usepackage{graphics}
\usepackage{subfigure}
\usepackage{pdflscape}
\usepackage{tabularx}
\usepackage{booktabs}

\renewcommand{\thispagestyle}[2]{}

\fancypagestyle{plain}{
        \fancyhead{}
        \fancyhead[C]{first page center header}
        \fancyfoot{}
        \fancyfoot[C]{first page center footer}
}
\begin{document}

%
\title{Software Defined Security Service Provisioning Framework for Internet of Things}

%
\author{Faraz Idris Khan, Sufian Hameed\\
       IT Security Labs, National University of Computer and Emerging Sciences (FAST-NUCES), Pakistan\\
       faraz.idris@nu.edu.pk, sufian.hameed@nu.edu.pk}

%


\maketitle

\begin{abstract}
	Programmable management framework have paved the way for managing devices in the network. Lately, emerging paradigm of Software Defined Networking (SDN) have revolutionized programmable networks. Designers of networking applications i.e. Internet of things (IoT) have started investigating potentials of SDN paradigm in improving network management. IoT envision interconnecting various embedded devices surrounding our environment with IP to enable internet connectivity. Unlike traditional network architectures, IoT  are characterized by constraint in resources and heterogeneous inter connectivity of wireless and wired medium. Therefore, unique challenges for managing IoT are raised which are discussed in this paper. Ubiquity of IoT have raised unique security challenges in IoT which is one of the aspect of management framework for IoT. In this paper, security threats and requirements are summarized in IoT extracted from the state of the art efforts in investigating security challenges of IoT. Also, SDN based security service provisioning framework for IoT is proposed.
\end{abstract}


\IEEEpeerreviewmaketitle

\section{Introduction}
Recent revolution in embedded technologies and Internet have made it possible for the things surrounding us to be interconnected with each other \cite{Atzori2010}. It is expected in the coming era IoT devices will be part of the environment around us which will generate enormous amount of data. Processing is required on the generated data which is then presented in an understandable form to the requester.

\begin{figure}
    \centering
    \includegraphics[width=1.0\linewidth]{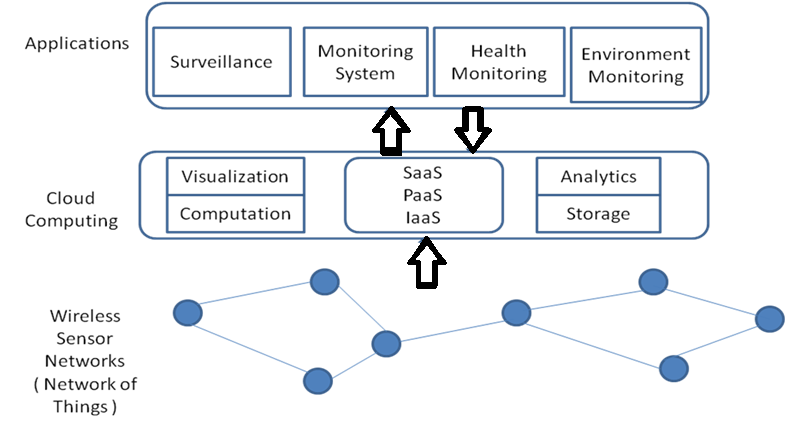}
    \caption{IoT and Cloud Convergence}
    \label{fig:Figure1}
\end{figure}

Mobile operators, software developers, integrators and alternative access technology are involved in the IoT ecosystem  \cite{Levakov2012}. There are many different application domains where IoT plays crucial role like manufacturing, health-care, transport, administration, insurance, public safety, local community, metering, road safety, traffic management, tracking, etc. IoT enables interconnection with people's devices exchanging information and performing actions with out humans involved. This is possible by amalgamating heterogeneous communication infrastructure. This has motivated the researcher to design smart gateways which connects IoT devices with traditional internet. Most recently, enabling \emph{Everything as a Service model} by merging IoT and Cloud Computing is the focus of attention in the research community \cite{gubbi2013internet} (see figure \ref{fig:Figure1}).

In order to tackle management problems in IoT, resource management frameworks have received considerable attention. SDN paradigm offers an attractive solution to manage IoT resources which is been lately under focus. Proposal of a framework for managing traffic and network resources in an IoT environment is given in \cite{HaiHuang2014}. Other efforts which have adopted SDN based approach to solve management issues in IoT can be found in \cite{Flauzac2015}\cite{Omnes2015}\cite{Qin2014}\cite{Hu2015}\cite{Jararweh2015a}\cite{Tadinada2014}.

Numerous security challenges are illuminated with the increasing complexity of IoT networks. There is a desire for a complete framework which manages data generated from the IoT devices. Up till now there is no SDN based comprehensive security framework for IoT network. For managing security of IoT networks the promise of SDN to manage IoT resources makes it a prime candidate for management framework. Contribution in this paper is twofold which are as follows.

\begin{enumerate}
  \item Identified and discussed management challenges of IoT. Furthermore in this paper security management of IoT is dealt with and security issues, threats, attacks and requirements in IoT are identified.
  \item Proposed SDN based framework for provisioning security services over IoT network.

\end{enumerate}

The paper is organized as follows. Section 2 gives a brief introduction on software defined networking paradigm. Management challenges of IoT are discussed in section 3. Section 4 gives detail insight into security issues and requirements of IoT. Section 5 discusses security threats and attacks in IoT. In section 6, general introduction to software defined networking framework for IoT (SDIoT) is given.  Section 7, presents the proposal of SDN based IoT framework for security service provisioning. In the end, conclusion and future work is discussed.

\section{Introduction to SDN}
\label{sec:sdn}
Developments in programmable networks by Martin Casado, Nick Mckeown and Scott Shenker at Stanford University resulted in a novel paradigm of SDN. Road to SDN with a brief discussion on the history of programmable networks can be found in \cite{Feamster2014}. SDN enables network administrators to manage network services by abstracting high level functionalities. Such abstraction is provided by separating control plane from the data plane. With such separation network management is simplified.

Appropriate forwarding decisions for the end network devices which are SDN enabled are given by a special central node called SDN controller. Openflow is the most widely used protocol used by the end devices to communicate with the SDN controller. Openflow supported devices are often called Openflow switches. An OpenFlow switch separates the data plane and control plane functions. High level routing decisions are moved to the controller. Switch have the data plane portion. Figure 2. illustrates a typical SDN architecture.

Secure connection with the network devices and SDN controller is established by Openflow control message exchanges. Furthermore, the message exchange result in installing forwarding instructions. Flow tables in the switches are maintained by the data plane. In flow tables, each flow table entry contains a set of packet fields to match an action (such as send-out-port, modify-field, or drop). Upon receiving a packet which is never seen by the Openflow switch and no matching flow entries is found. The switch sends such packet to the controller for making decision. Decision taken by the SDN controller can be of dropping the packet or adding flow entry in the Openflow switch for forwarding in future. Flow tables have flow entries which are defined by flow rules defined in Openflow. Heterogeneous network changes result in dynamic modification of flow rules.

\begin{figure}
	\centering
	\includegraphics[width=1.0\linewidth]{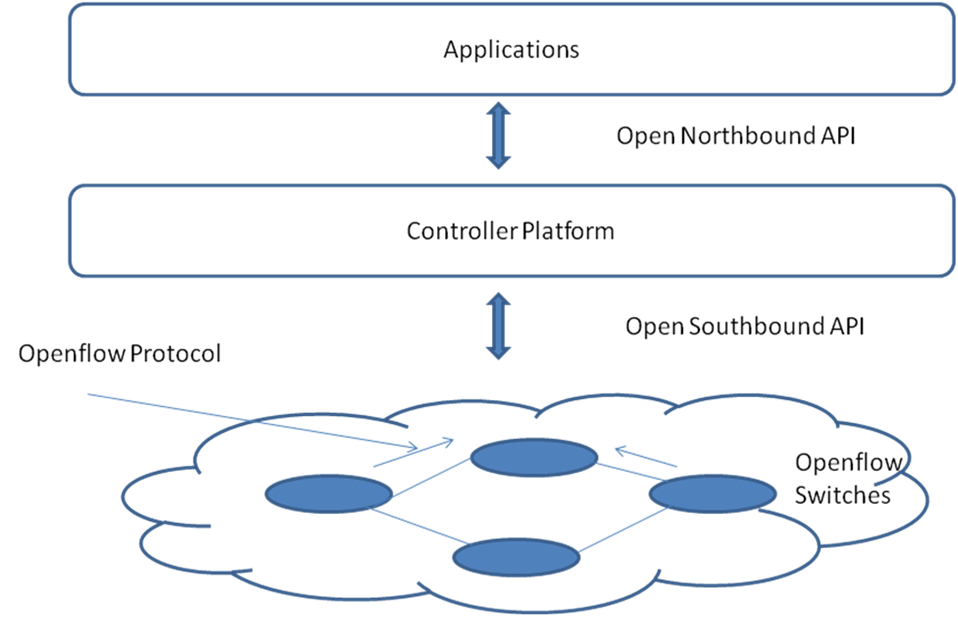}
	\caption{SDN Architecture}
	\label{fig:Figure2}
\end{figure}

\section{Management Challenges of IoT}
\begin{figure}
	\centering
	\includegraphics[width=1.0\linewidth]{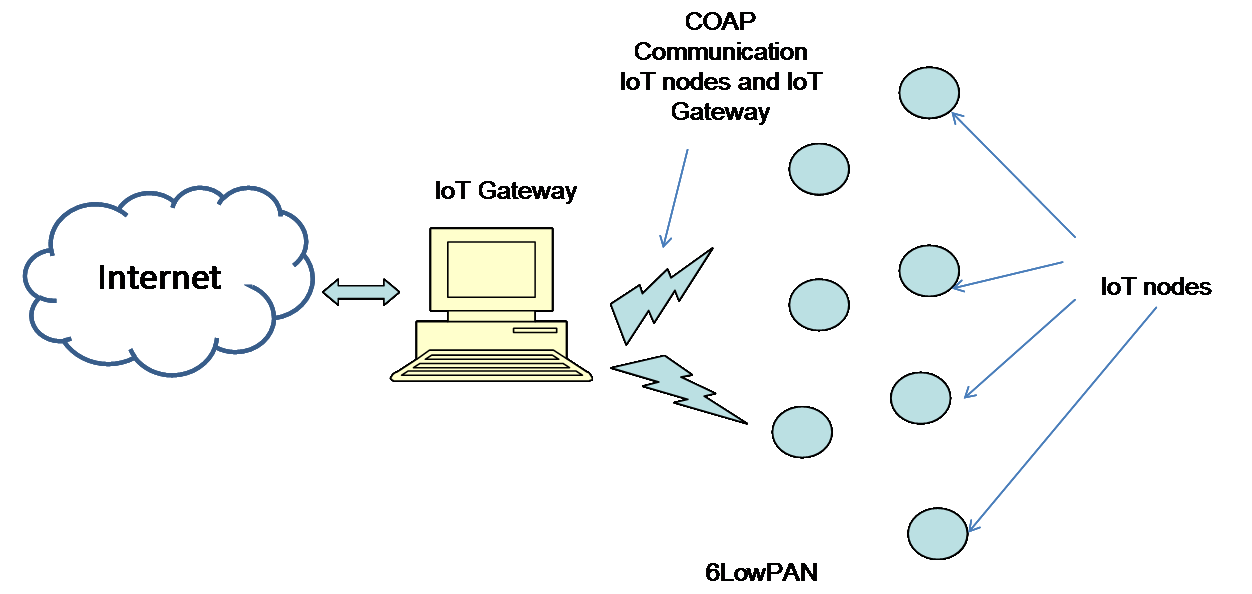}
	\caption{IoT devices connected with Internet via IoT Gateway.}
	\label{fig:Figure3}
\end{figure}
Conventional network management techniques are inapplicable in IoT due to distinctive challenges. IoT devices connected to the internet via gateway is shown in figure 3. CoAP ( Constrained Application) protocol running over 6LowPAN is used for communication between gateway and IoT nodes. IoT network devices are not sufficient in resources. Usually in IoT network high fault rates are experienced due to shortfall in energy and connectivity interruptions. To improve the efficiency of the network main concerns are of monitoring and administration of node communication. A typical management solution should provide various management functions integrating configuration, security operation, administration of devices and services of IoT. Following set of functions should be provided by management solution for IoT.

\subsection{Fault Tolerance}
Failures are encountered in the IoT network for various reasons. Depletion of batteries is one possibility. Effects on the sensing components results in inaccurate readings disseminated by the devices. Vigorous changes in network topology and partitions in the network is created due to inherent nature of adhoc wireless network links tendency of failures. Due to erroneous nature of communication, delivered packets get corrupted. Packet losses are not experienced due to failures of link but are also caused by congestion. Multihop communication nature of IoT worsen all the fault scenarios discussed. Efforts in this direction is summarized in \cite{Paradis2007a}.

\subsection{Energy Management}
IoT network are deployed in distant region. Due to scarcity of energy resource in IoT and its deployment in distant region depletion of available energy happens frequently. Substitute of energy is impossible. Balanced energy management among supply and load is required to avoid energy scarcity. Data traffic from devices can be controlled to balance energy in the network which is possible by techniques such as duty cycling, scheduling sleep and wake-up modes of devices studied in \cite{Khana}. Management solution for IoT should address the energy issue by having essential function of energy management for smooth operation of IoT network.

\subsection{Load balancing}
Load balancing can be used for extending lifetime of IoT which results in lessening energy utilization. Load balancing is possible by techniques such as clustering. In clustering technique IoT network is organized into cluster which is coordinated by a cluster head. With such configuration there are numerous benefits, such as reduction in routing table size, conservation of network bandwidth, lengthening network lifetime, reduction in redundant data packets and decreasing energy consumption. This makes load balancing an essential component of management solution for IoT. Efforts in this direction can be found in \cite{Wajgi2012}.

\subsection{Security Management}
Security, privacy and trust are essential requirements in IoT. Due to resource constraint nature of IoT devices the the provisioning of security has become more challenging. Novel techniques for provisioning of security is required as conventional security schemes are inapplicable. This poses unique challenges for management framework designer of IoT. Research on security management can be found in \cite{Ali2013}.

In this paper, security management issues of IoT network are dealt with.In addition, major security issues and requirements put forth by IoT are identified. In order to address these issues and requirements SDN based framework to provision security services over IoT is proposed.

\section{Security Issues and Requirements in IoT}

In this section, major security issues and requirements are identified which are imposed by resource constrained IoT network. Traditional solutions are inapplicable in the domain of IoT due to resource constraint nature of IoT devices. In this section latest research efforts in each of the issues and requirements are outlined.

\subsection{Privacy in IoT}
IoT finds its uses in various heterogeneous fields which include remote monitoring of patients, control of energy consumption, traffic control, production chain, smart shopping etc. Hence, privacy in IoT is a prime security issue which needs full attention from researchers in academia and industry. There is a desire need to propose protocols and management framework for handling privacy in IoT. Latest attempts to address the issue can be found in literature such as \cite{Sciancalepore2015}, \cite{Liu2015}, \cite{Journal2015}.

\subsection{Lightweight Cryptographic Framework for IoT}
IoT is usually scarce in resources and faces number of challenges such as limitation in power and bandwidth, .etc. Hence traditional heavy weight algorithms cannot be opted for IoT. There is a desire need to investigate how to make symmetric and asymmetric algorithms lightweight for IoT. Most recently, various efforts for lightweight solution for IoT can be found in \cite{Lakkundit2015}, \cite{Hernandez-Ramos2015a}, \cite{Lakkundit2015}, \cite{Lakkundit2015a}. Some of the existing lightweight cryptographic solutions includes HIGHT, RC5 and PRESENT.

\subsection{Secure Routing and Forwarding in IoT}
IoT not only require provisioning of security services but often experience problems in routing and forwarding the data. Hence, there is a desire need to secure the routing algorithm for IoT. IoT routing usually experiences selective forwarding attacks, sink hole attack, Hello flood attack, Wormhole attack, clone ID and sybil attacks. Efforts have been putforward by the researchers to address routing attacks in IoT. A comprehensive state of the art in securing routing for WSN can be found in \cite{MdZin2014}. Latest efforts in proposing secure routing for IoT can be found in \cite{Cervantes2015}, \cite{Duan2014}, \cite{Tang2015}.

\subsection{Robustness and Resilient Management in IoT}
IoT applications require robustness as failures due to security attacks may not be affordable for low power devices. Hence, an ideal IoT management framework should have mechanisms to prevent such situations and ensure fault tolerance against security failures. Efforts in this direction can be found in \cite{Gia2015a}, \cite{Garcia-Morchon2013a}, \cite{Kamal2014}, \cite{Sommer2014}.

\subsection{Management Framework for IoT}

SDN offers tremendous amount of opportunities to manage future Internet. The conventional network protocols and equipment is not able to support huge traffic amount, mobility and elevated level of scalability. In literature, authors have proposed architecture for IoT \cite{Diogo2014}. Most of the researchers who have realized the potential of SDN for managing IoT network have proposed architecture which apply SDN and network virtualization for managing IoT networks \cite{Diogo2014}.  Interestingly, authors uptill now have discussed at a very primitive level to provide security for IoT using SDN. Some of the latest work in this direction can be found in \cite{Olivier2015}, \cite{Omnes2015}, \cite{Hu2015}, \cite{Choi2016}. A complete security framework to provide security in industrial IoT has received very little attention.

\subsection{Access Control in IoT}

Access control allows only authorized users to access the resources. In any typical IoT scenario like Smart Home, lack of proper access control mechanisms could result in disclosing sensitive and privileged information. It is very important that the user data is only disclosed to authorized parties.

\subsection{Trust in IoT}

In safety critical IoT applications (like health care), the trustworthiness of sensors and sensor's data is important. Any malicious sensor node or malicious sensor data can lead to a disaster. It is important to calculate trust and reputation of different entities involved in the IoT ecosystem.


\subsection{Audit Control for IoT}

IoT environment needs to know when their services are accessed and by whom. This will ensure in maintaining higher level of security. Maintaining an audit trail in IoT services is a challenging task. Having a centralized view of IoT network with SDN managed framework can help in logging activities across IoT network.

\subsection{Secure Network Access}
IoT devices when joining the network need to be authorized in order to avail services from the network. In case, a malicious node becomes part of the network it can perform malicious actions which either cause disruption in IoT services or modify the critical data with in the IoT network. There is a desire need to authorize IoT nodes entering the network with authentication algorithms and enforce security policies in order to inhibit IoT nodes to perform unauthorize actions. Such issues have been looked upon and discussed in literature such as in \cite{Mordeno2015}.

\subsection{Secure Storage}
Recent advances in flash memory storage technology have enabled IoT devices such as smart objects to have enormous amount of storage space. Most recently, IoT applications store data in IoT devices for improved performance \cite{Tsiftes2011}. Protecting sensitive data information have received attention from the research community. Stored data in IoT devices can be tampered and modified by malicious nodes or applications with in the network. Efforts in this domain can be found in \cite{Bagci2012} and \cite{Tsiftes2009}.

\subsection{Tamper Resistance}
Malicious entities within the network can take hold of the IoT device or devices which can be tampered. Hence, an IoT device should be resistant to such tampering and fulfills all the security requirements. In literature, need for resistance against tampering is an issue that need to be addressed \cite{Babar2011}. Hence, robust security management solution is desired which ensures resilience against such tampering.

\subsection{User Identification and Identity Management}
A node joining the network have to be authenticated. Along with authentication it is to be made sure that joining node is validated with the policies applicable to the particular joining node. It is desire to maintain ID of all the nodes in the network. Identity management is required for IoT and effort in this direction can be found in \cite{Chibelushi2013}.

\subsection{Availability}
IoT devices usually sense data which is collected, aggregated and used by IoT application. IoT applications apply data analytics to infer from the collected data which is then used for making decisions. It is possible that attacks can be launch on IoT devices which hinders their proper operation and availability leading to unnecessary delays and errors in analyzing data. Availability can be targeted by (Denial of Service) DoS or (Distributed Denial of Service) DDoS attacks on IoT nodes or network. DoS and DDoS have received considerable attention from the research community and notable efforts can be found in \cite{Kasinathan2013b}.

\section{Security Threats and Attacks in IoT}
\label{sec:threat}

This section lists the security threats and attacks, which are applicable to IoT ecosystem. Efforts in summarizing challenges of security in IoT can be found in \cite{Singh2015} \cite{Babar2010} \cite{Pimentel2012} \cite{Suo2012} \cite{Heer2011a} \cite{Li2016}. Detail taxonomy of security challenges can be found in \cite{Babar2010}.On the other hand, comprehensive identification of security challenges and requirements in IoT architecture is also carried out by various researchers most notable of them can be found in \cite{Suo2012} \cite{Heer2011a} \cite{Li2016} \cite{Singh2015}. Challenges from all the above efforts are summarized which are required to be addressed by a comprehensive security management solution for IoT. These threats and challenges are given below.

\subsection{Eavesdropping}

Because IoT involves wireless communication interface it is obviously vulnerable to eavesdropping. IoT services are expected to contain sensitive data, therefore it is important to protect the data of IoT connected objects against an eavesdropper for possible data leakages. Consider a Smart home environment where IoT objects control and monitor different activities. It is of major importance that the personal information of the smart home owner is kept private and the attacker/eavesdropper is not able to tap the communication between IoT devices.

\subsection{Data Corruption}

Instead of eavesdropping or listening to the data, an attacker can also try to modify the data which is transmitted over the air between the IoT devices. A simple motivation here would be to disturb the communication such that the receiving device is not able to understand and process the data sent by the other device. This attack is a simple form of DoS attack where the devices are not been able to perform the required operation over the data. Other then that this attack does not allow the attacker to manipulate the actual data.

\subsection{Data Modification}

IoT nodes are expected to exchange critical data with other services and some time also with intermediate entities i.e. authorities, service providers and control centers. This put stringent requirement that the sensed, stored and transmitted data must not be tampered either maliciously or accidentally. In data modification the attacker is capable of manipulating/modifying the data in such a way that the receiving device is unable to detect modification and treats the input to be valid. This is very different and sophisticated from just data corruption attack. It is crucial to design reliable and dependable IoT applications secure against active modification.

\subsection{Identity Spoofing Attack}
IoT device's identity can be compromised through which malicious traffic is sent to victim nodes in the network. This is a devastating attack which can disrupt the normal operation of IoT network. This can also be used to launch DoS and DDoS attacks in the network. There is a strong desire for a robust and resilient techniques for validating IoT devices in the network to prevent spoofing.

\subsection{Injection Attack}
IoT devices run lightweight code to assist IoT applications in sensing, data collection or performing some activity in a particular region in a field. There is a possibility a malicious code can be injected by the attacker on IoT devices which then perform malicious actions with the intention of disrupting normal operation of IoT network or application. Malicious code can also sabotage IoT device in the network.

\subsection{Denial of Service and Insider Detection in IoT}

Emerging technology of IoT experiences severe attacks in IoT. Insider attack is one of the devastating attacks that has received attention from researchers \cite{Hu2014}. It is desired to address the issue by incorporating a mechanism in security management framework for IoT.





\subsection{Attacks on Availability}


Availability is extremely important for IoT services which enable access from anywhere at any time in order to provide information continuously. Existing security protocols fails to effectively prevent attacks on availability of IoT services. Lets consider an example of Smart Home application, where the sensor nodes are incapable of handling huge number of requests due to resource limitations. Attacker can leverage this limitation to launch DoS attack by sending huge volume of false service requests. Since the wireless transmissions are also battery hungry operation, unnecessary handling of service request will also drain the battery of IoT devices. Security management framework for IoT should be able to mitigate DoS attacks in IoT. Efforts in dealing with DoS attacks in IoT can be found in \cite{Kasinathan2013a}, \cite{Nigam2014}.


\subsection{Impersonation Attacks}

In an IoT ecosystem, both the service provider and service consumer need to make sure that the service is accessed by the authenticated users and it is also offered by an authentic source. It is very crucial to have strong authentication mechanisms deployed to prevent any form of impersonation.

\section{Software Defined Network based Framework for IoT}


\begin{figure}
	\centering
	\includegraphics[width=1.0\linewidth]{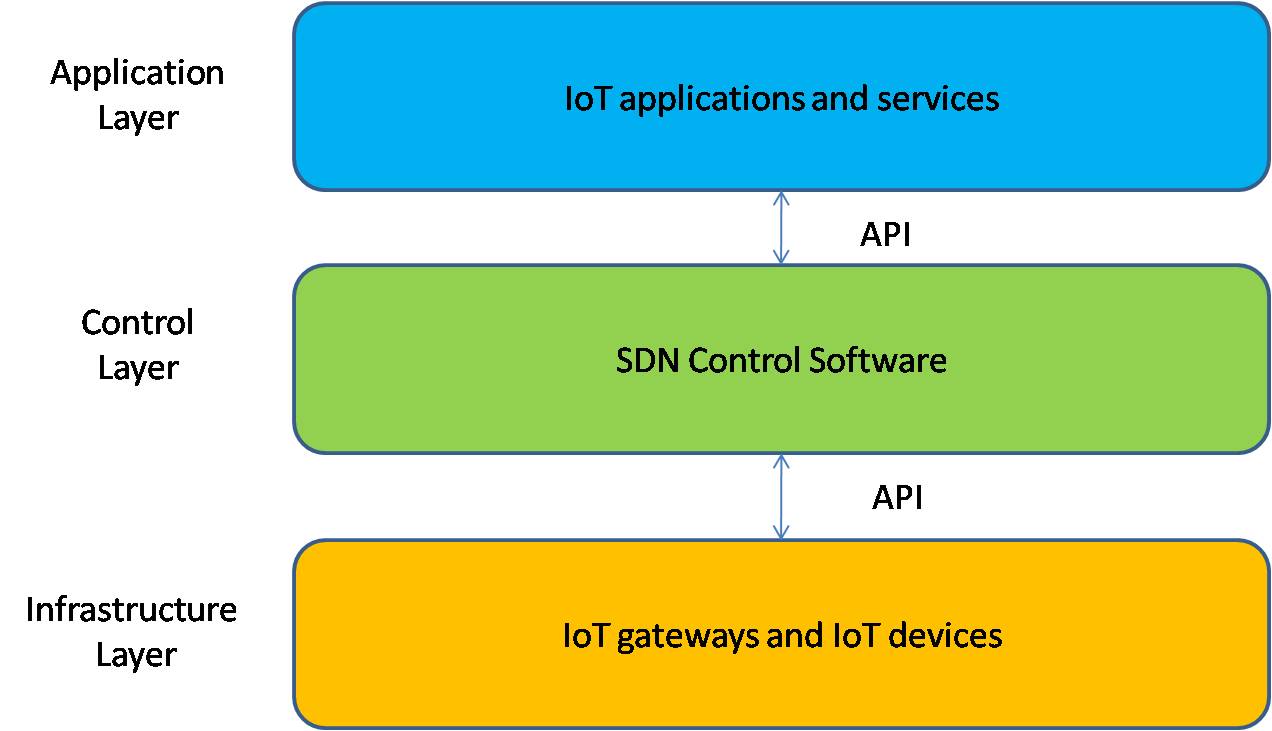}
	\caption{SDN Archtiecture for IoT.}
	\label{fig:Figure4}
\end{figure}

The heterogeneity of IoT multi-networks and its complexity is a challenge to organize and to make effective the use of heterogeneous resources with the objective of managing and securing as numerous jobs as possible. The researchers have considered SDN a hot candidate for solving resource management needs of IoT environment. This is due to inherent nature of SDN paradigm which is managed by a centralized controlling agent i.e. controller. Current practical implementation of SDN technologies are long way dealing with diversified and vigorous demands of IoT multi-network. Although, there are various differing SDN based solution for IoT a generic architecture can be constructed from the existing solutions in the literature as shown in Figure 4. In this architecture, IoT applications and services are implemented at application layer. SDN controller related functionalities is implemented at control layer. While IoT devices and gateway exists at infrastructure layer. The control software interacts with the IoT application services at application layer and IoT devices at infrastructure layer using APIs. Various efforts in SDN based solutions for IoT can be found in literature some of the important ones are \cite{Jararweh2015}, \cite{Omnes2015}, \cite{Hu2015}, \cite{Li2015} and \cite{Qin2014}.

\subsubsection{Sensor Openflow Switch(es)}

The IoT nodes are usually laid out in clusters with cluster head, which is a resource sufficient device that communicates with the IoT gateway. In order to implement SDN techniques the IoT nodes acting as relays or switching device play the role of sensor openflow switches. In contrast to traditional network components over the Internet, the IoT nodes are constrained in resources and require a lightweight openflow protocol for communication with the low power devices. The detailed design for the lightweight Openflow in IoT nodes requires resource efficient approach.

\begin{figure*}
	\centering
	\includegraphics[width=0.6\linewidth]{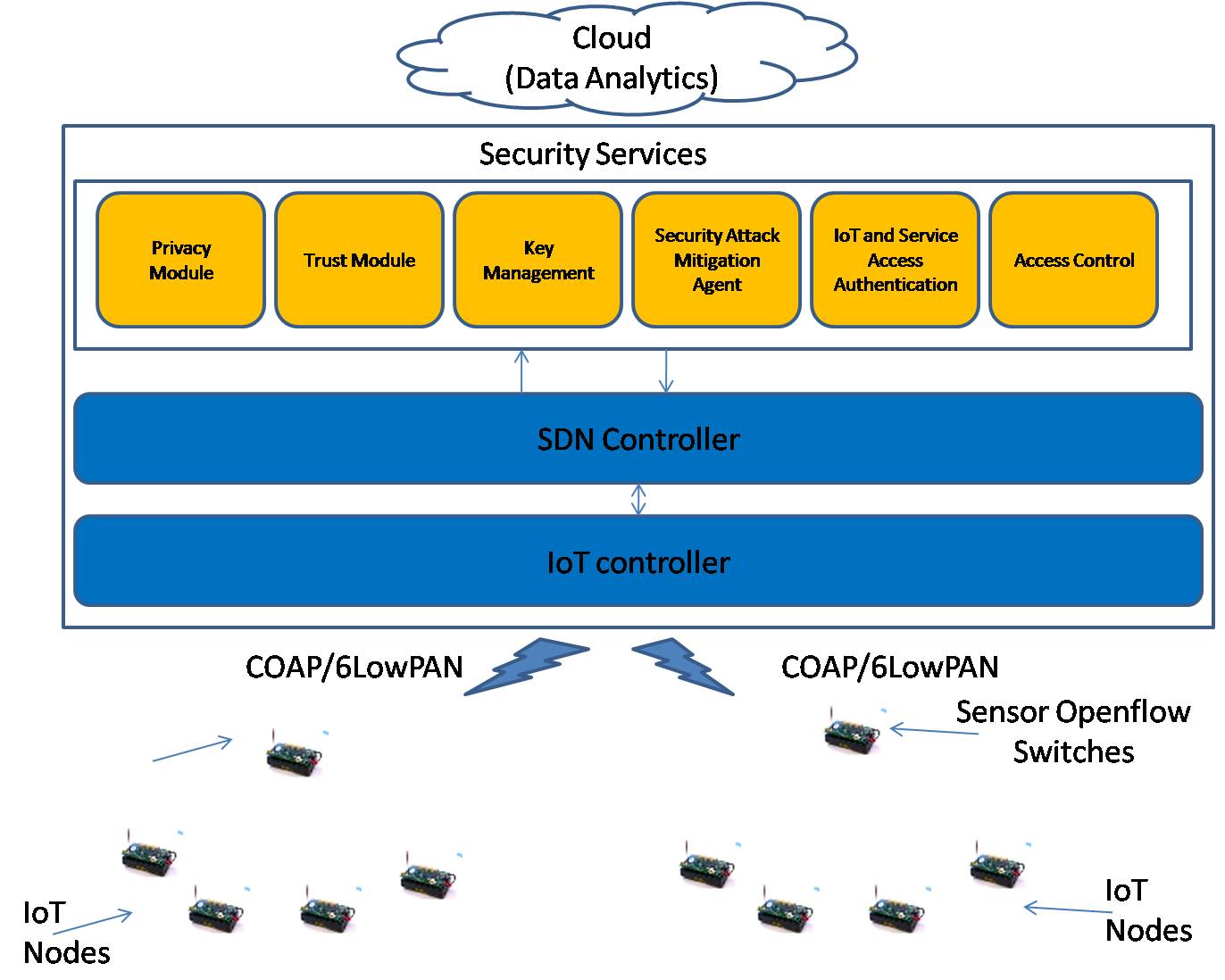}
	\caption{Proposed SDN based Management and Security Framework for IoT}
	\label{fig:Figure5}
\end{figure*}

\section{SDN based IoT Framework for Security Service Provisioning}
\label{sec:design}

Current architecture of IoT contains sensor nodes (IoT devices) connected to the conventional Internet via the IoT gateway. This architecture (IP connected IoT) is increasingly becoming popular today, where smart devices also referred to as the things are integrated with the Internet to form IoT.  Low power Wireless Personal Area Network running with IPv6 (6LoWPAN) is the popular technology used as a communication technology.

SDN makes network services agile and flexible such that they can be automatically deployed and programmed. It further simplifies network management by separating the control plane (network intelligence that make data forwarding decisions) from data plane (forwarding elements). The control plane functions are moved into central Controller which acts as the brain regulating the whole paradigm.

In order to address the issues discussed in this paper SDN based framework for providing security services to IoT network is proposed. The framework consists of an IoT controller and SDN based security controller (see figure. 5). Both of these controllers are located in IoT gateway, which communicates with the IoT devices. Most commonly used topology by IoT network is cluster based topology. Where cluster head manages a cluster of IoT devices.

The proposed SDN based IoT framework essentially comprise of three main components.

\begin{enumerate}
	\item IoT Controller.
	\item SDN based Security Controller.
	\item Sensor Openflow Switch(es) (briefly discussed above).
\end{enumerate}

\subsection{IoT Controller}

The IoT controller act as a middle tier collecting information from IoT devices and transmitting it to application services for data analytics. It is responsible for data collection, aggregation and transmission of data to the back end. This is realized through a monitoring agent that collects data across the IoT network as shown in figure 5.

\subsection{SDN based Security Controller}

The SDN based security controller is also placed in the IoT gateway and run on top of the IoT controller. In order to realize security provisioning with the IoT network the SDN based security controller interacts with the IoT controller to monitor the flows. The security controller utilize SDN techniques to provide different security services across IoT network. The SDN based controller interacts with security applications at the application plane to provision i) Privacy ii) Trust iii) Key Management iv) Access Control v) Service Access Authentication across IoT network and vi) Security Attack Mitigation. Figure 5. shows the SDN based framework for provisioning security where the security services are implemented at application plane of the SDN architecture. The network administrator will enforce security policies through the security applications by using custom API.

Security services at application plane will require status of the network nodes in IoT. Flow samples required by the network application is  given by SDN based controller at the control plane which has the whole global view of the network. SDN based IoT controller have services which are implemented as modules to provision security services across IoT network. These modules are as follows.

\subsection{Privacy Module}

\begin{figure}
	\centering
    \includegraphics[width=0.6\linewidth]{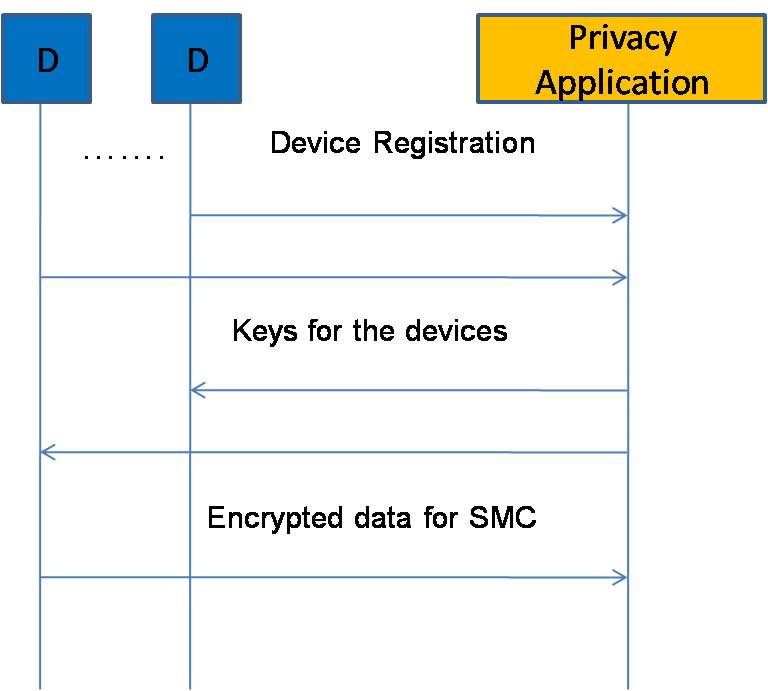}
	\caption{Privacy process}
	\label{fig:Figure 6}
\end{figure}

\begin{figure}
	\centering
    \includegraphics[width=0.5\linewidth]{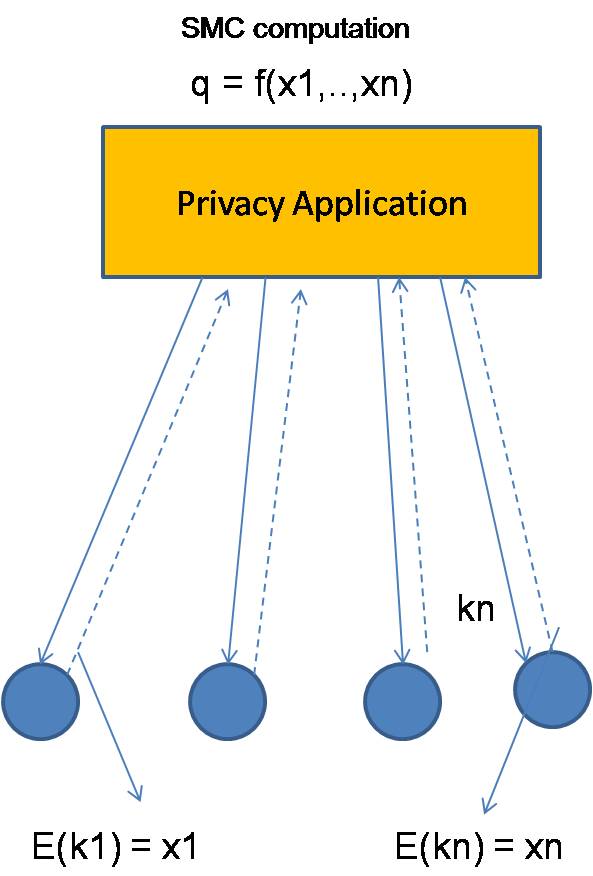}
	\caption{SMC Computations}
	\label{fig:Figure 7}
\end{figure}

\begin{figure}
	\centering
    \includegraphics[width=1.0\linewidth]{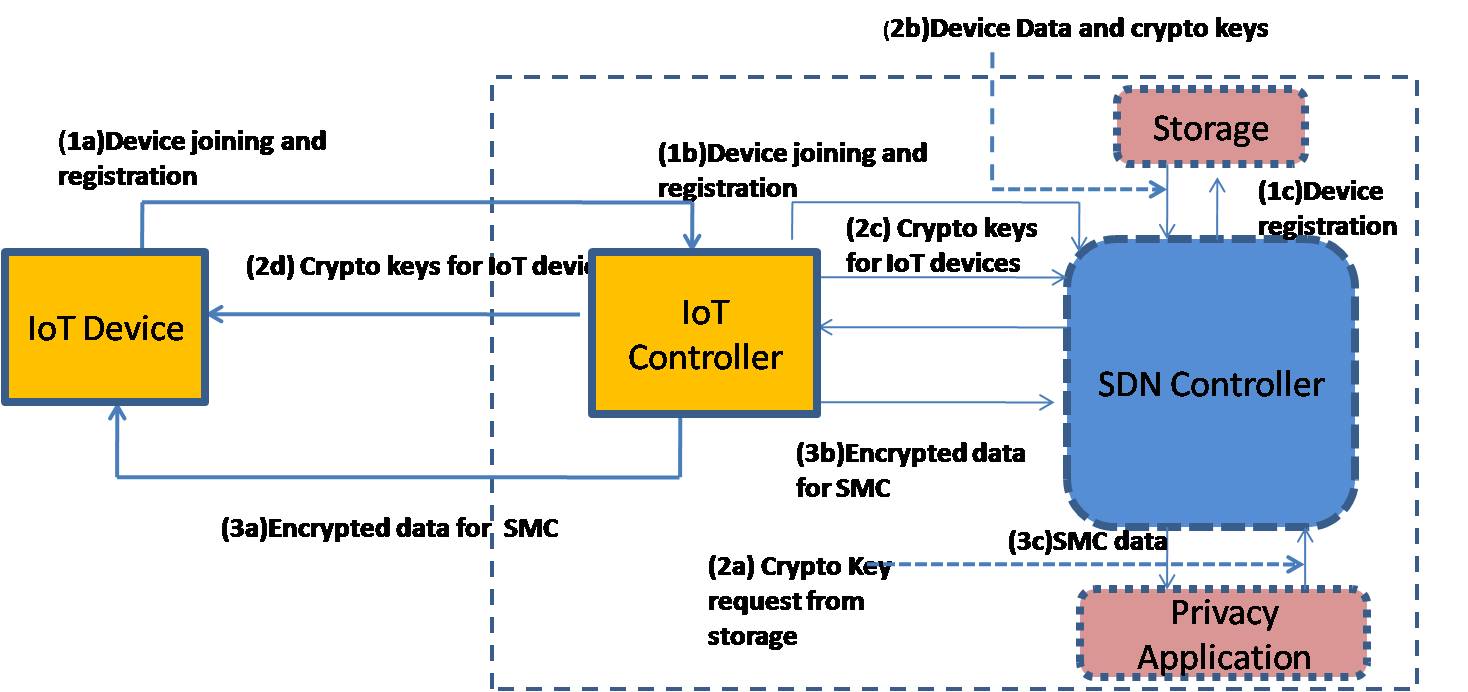}
	\caption{Privacy Module}
	\label{fig:Figure 8}
\end{figure}

This component will preserve privacy in the IoT network. Privacy means that the data generated by the devices need to be transmitted anonymously without revealing any information this to the intermediate or unauthorized nodes attempting to eaves drop on data. Generated data from IoT devices are collected and aggregated at the IoT controller. Privacy in the process of data collection and aggregation has to be ensured in IoT network which can be compromised by adversary either at collection or aggregation. SDN controller enables flows from each device on the IoT controller to centrally verify the authenticity of data originated from a device. This is possible by public key infrastructure (PKI) when the device is joined and registered the cryptographic credential valid during the lifetime of device is given.

Feasibility of PKI in low power devices has been studied a lot by the research community. It is often argued to be computationally intensive and practically not possible in IoT. Lately, making PKI feasible for IoT is under focus by the research community \cite{Roman2007}\cite{Computing2005} and ECC ( Elliptic Curve Cryptography) is being employed to adopt PKI in IoT \cite{Malan2004}. Originated data from each device is encrypted by public key of the gateway which is then signed by the private key of the device. Encrypted packets in the flows from devices in the IoT network are then verified through SDN controller by the privacy application. In order to preserve privacy of the aggregated data Secure Multiparty Computation (SMC) \cite{Clifton2002} is incorporated which originated from the work of generating and exchanging secrets among two parties \cite{Yao1986}. SMC is used in preserving privacy \cite{Lindell2009}. The principle of SMC works in a manner in which computation is secure when no party have the knowledge of anything except for the input and the result. Mathematically, given inputs \[ (n^1,n^2,...n^n) \] from the sources \[(1,2,3,...n)\] which are to be processed by a trusted intermediate component as \[f(n^1,n^2,...n^n)=y\] and then announce the results. Intermediate trusted component is a trusted third party (TTP) which keep the process anonymous by aggregating inputs from the devices. In our model, SDN controller along with privacy preserving application acts as a TTP to perform SMC and preserve privacy of the whole data collection process.
Figure 8. shows the outline of the privacy module.

Privacy preserving process as shown in figure 6 and 7. is as follows

\textbf{Step 1:} IoT devices join the network by sending their request to the gateway which is collected by the IoT controller and passed on to the SDN controller which registers the device.

\textbf{Step 2:} Cryptographic credentials for confidentiality, integrity and SMC is generated by the privacy application which is stored for a registering device.

\textbf{Step 3:} Encrypted packets are sent from a device as a flow which is accounted by the SDN controller and along with privacy application verifies it and runs SMC algorithm. The computed result over inputs from IoT device is then provided to the back end application or data analytics possibly hosted in a cloud or if needed to the IoT devices for further processing.

\subsection{Trust Module}

\begin{figure}
	\centering
    \includegraphics[width=1.0\linewidth]{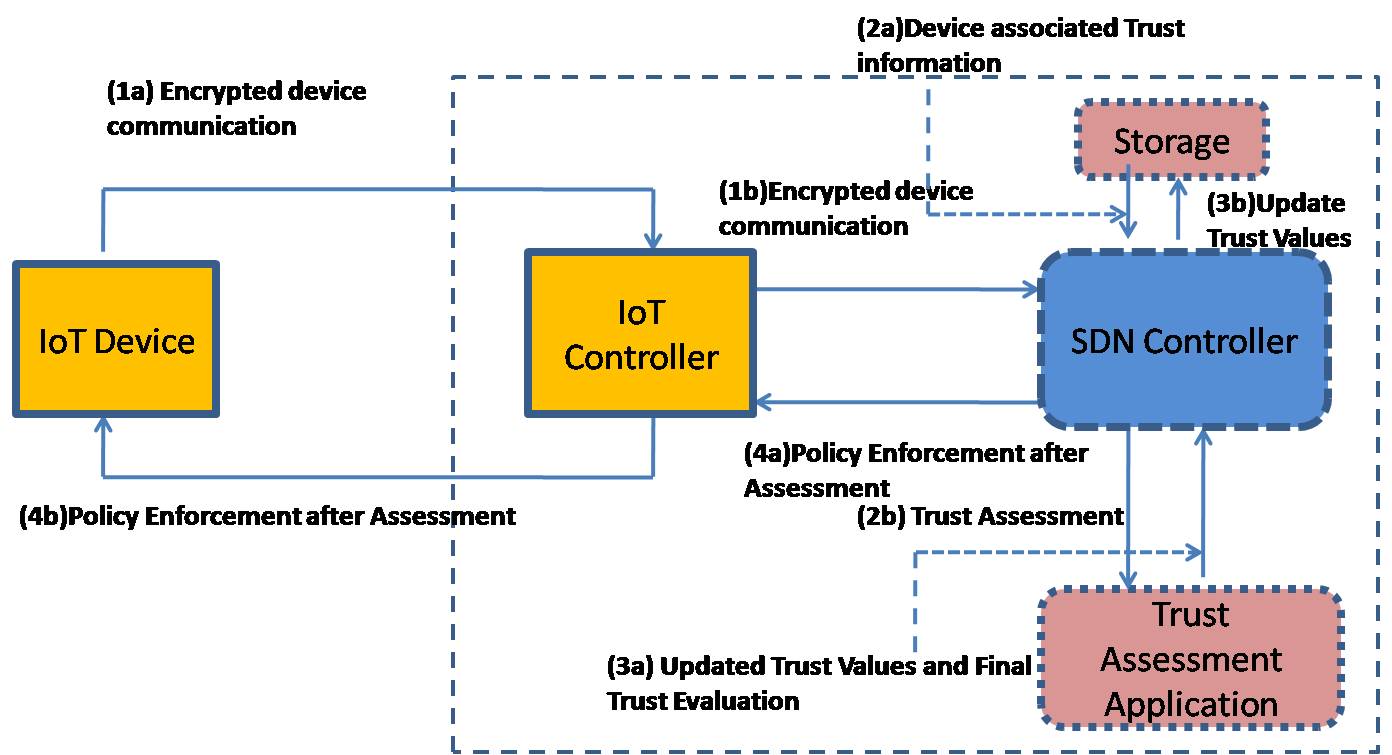}
	\caption{Trust Module}
	\label{fig:Figure 9}
\end{figure}

\begin{figure}
	\centering
    \includegraphics[width=1.0\linewidth]{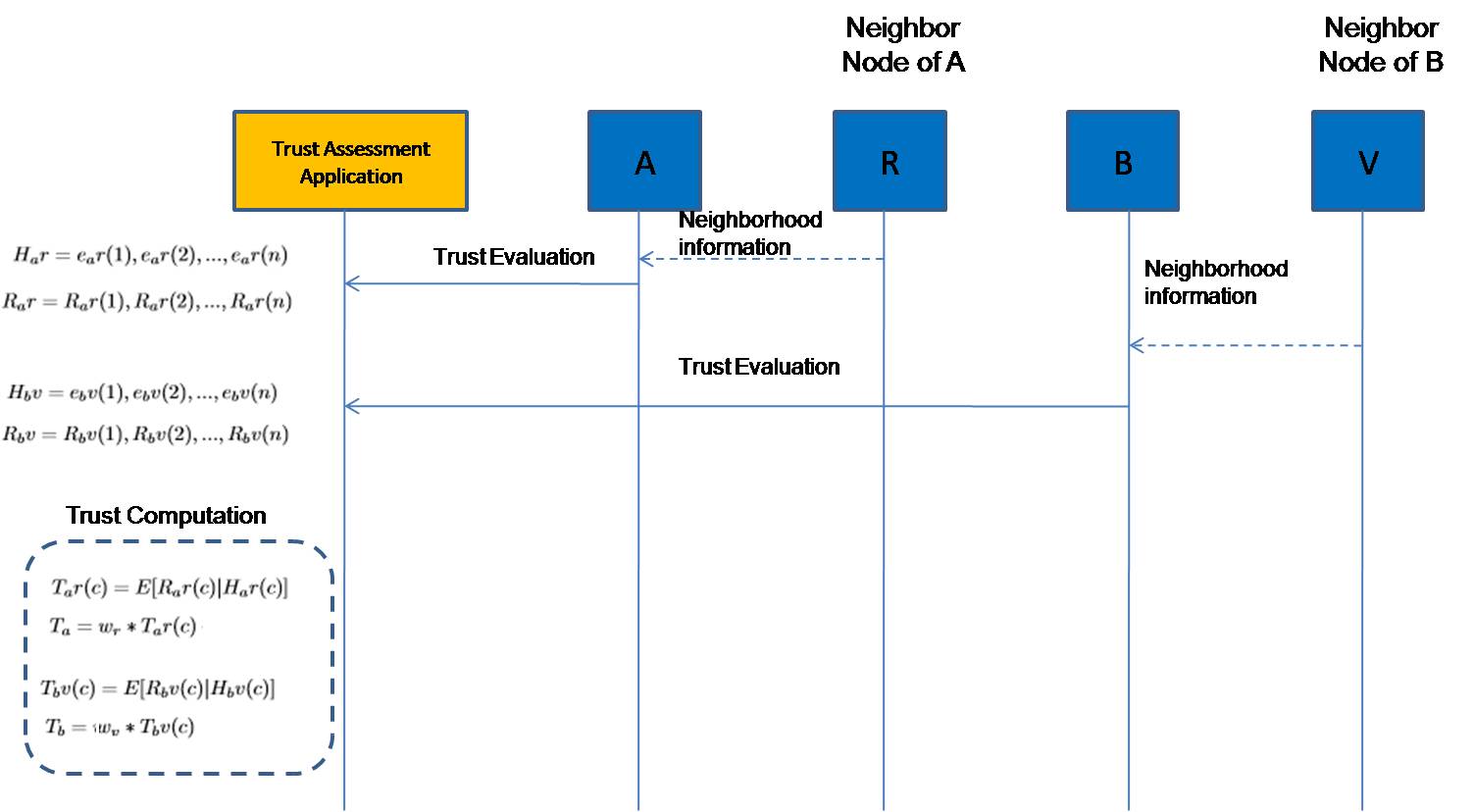}
	\caption{Trust Evaluation Process}
	\label{fig:Figure 10}
\end{figure}

Statistics of the IoT flows will be used to generate trust values across the IoT network. These trust values will be maintained by keeping a moving average over the historical statistics of IoT flows. An untrusted IoT node or flow attempting to manipulate the network will fail the trust evaluation test. In an IoT network there will be service requester and service provider. Service requester can be a thing or a user requesting a certain service. Service provider will first calculate the trust values of the nodes in collaboration with neighboring nodes. Trust values are computed on the basis of history maintained by keeping account of previous interactions of a device who's trust is to be computed with the neighboring device.  Trust assessment application is inspired by \cite{Mui2002} and \cite{Reddy2011}. For simplicity let us define history, if a binary random variable \[ (e_ab(i)) \] indicates an ith encounter between node \emph{a} and \emph{b}. \[ (e_ab(i)) \] can assume value of 1 if \emph{b}'s action is cooperate and 0 otherwise. Then history H defined for a set of n encounters between \emph{a} and \emph{b} is represented by:

\[H_ab={e_ab(1),e_ab(2),...,e_ab(n)}\]

Lets represent reputation of node between \emph{a} and \emph{b} for an ith encounter as follows:

\[R_ab(i) \in [0,1]\]

As node \emph{a} interacts with \emph{b} the quantity \[R_ab(i)\] is updated with time as \emph{a}'s perception with \emph{b} changes

Trust value is then given as
\[T(i)=E[R(i)|H(i)]\]

Higher the trust value for a particular node in the network higher the expectation that particular node will reciprocate with the other entity involved in the communication.

Lets assume that a particular node E who's trust is to be computed have neighboring nodes A, B, C and D who's weights are given as \[[w_a,w_b,w_c,w_d]\]. Then the trust of E is given as:

\[T(E)=w_a*T(A)+w_b*T(B)+w_c*T(C)+w*dT(D)\]

Hence, trust is computed by taking into account the interactions of the node who's trust is to be computed with the neighboring nodes in the network.

Figure 9. shows the outline of trust module. Trust Module process is as follows.

\textbf{Step 1:} The request from each node is sent as encrypted to the gateway. Request is collected by the IoT controller which is then relayed to SDN controller. SDN controller accounts flows in the network.

\textbf{Step 2:} Trust assessment for each request from either the thing or service is carried out by trust application. At this step historical data of trust evaluation is taken as an input to assess new trust values.

\textbf{Step 3:} Updated trust values are used to make decision about the policies to be enforced based on new trust computation. These policies apply to both devices and service.

\textbf{Step 4:} Based on trust assessment on updated values computed by the trust application the policies generated for the device may be sent to the device via IoT controller. Policy ensures that no device in future can perform illegal operations or access.

Trust assessment application process is as follows. Figure 10. shows how the process works.

\textbf{Step 1:} Upon receiving the request for trust evaluation, nodes involve in servicing the request are assessed in case an IoT service application interacts with the devices. For node to node communication historical interaction between the two nodes are assessed for trust evaluation.

\textbf{Step 2:} Lets say group of nodes A, B are involved for servicing request from the IoT service i then historical interactions and reputations of the neighboring nodes with A, B are evaluated. For simplicity it is assumed A have neighbor node R , B have neighbor node V. Historical interactions for 1,...,n and reputation are given as \[H_ar={e_ar(1),e_ar(2),...,e_ar(n)}\], \[H_bv={e_bv(1),e_bv(2),...,e_bv(n)}\],\[R_ar={R_ar(1),R_ar(2),...,R_ar(n)}\], \[R_bv={R_bv(1),R_bv(2),...,R_bv(n)}\].

\textbf{Step 3:} Trust value for neighboring nodes with the nodes involved in communication A, B (at an instance \emph{x} lying somewhere between 1, ..., n historical interactions) is then given as  \[T_ar(x)=E[R_ar(x)|H_ar(x)]\],\[T_bv(x)=E[R_bv(x)|H_bv(x)]\]. Trust for nodes A and B with given weights of neighboring nodes R and V \[w_r,w_v\] is given as \[T_a=w_r*T_ar(x)\], \[T_b=w_v*T_bv(x)\].

\textbf{Step 4:} Service i then assess the trust values of nodes A, B and proceed with its communication.

\subsection{Key Management}

\begin{figure}
	\centering
    \includegraphics[width=1.0\linewidth]{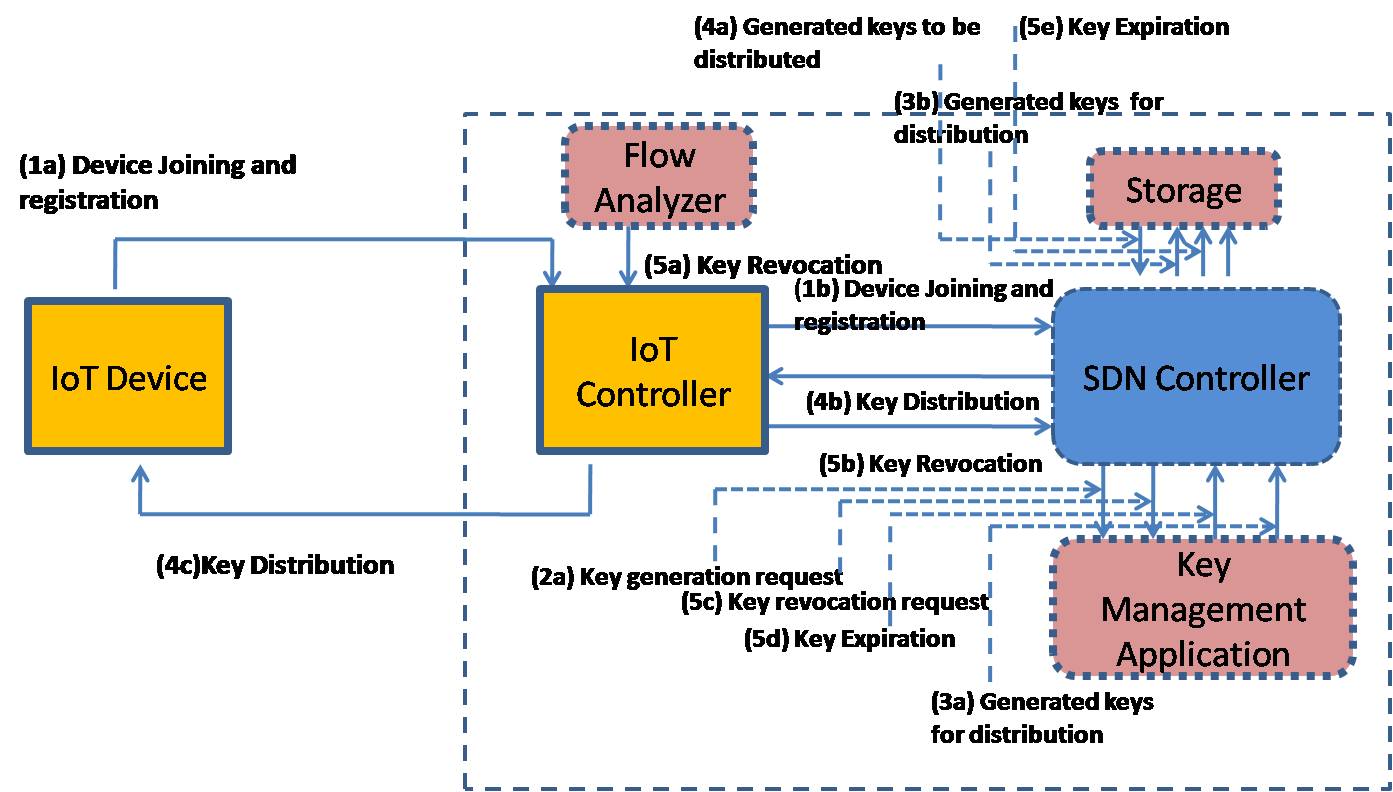}
	\caption{Key Management Module}
	\label{fig:Figure11}
\end{figure}

\begin{figure}
	\centering
    \includegraphics[width=1.0\linewidth]{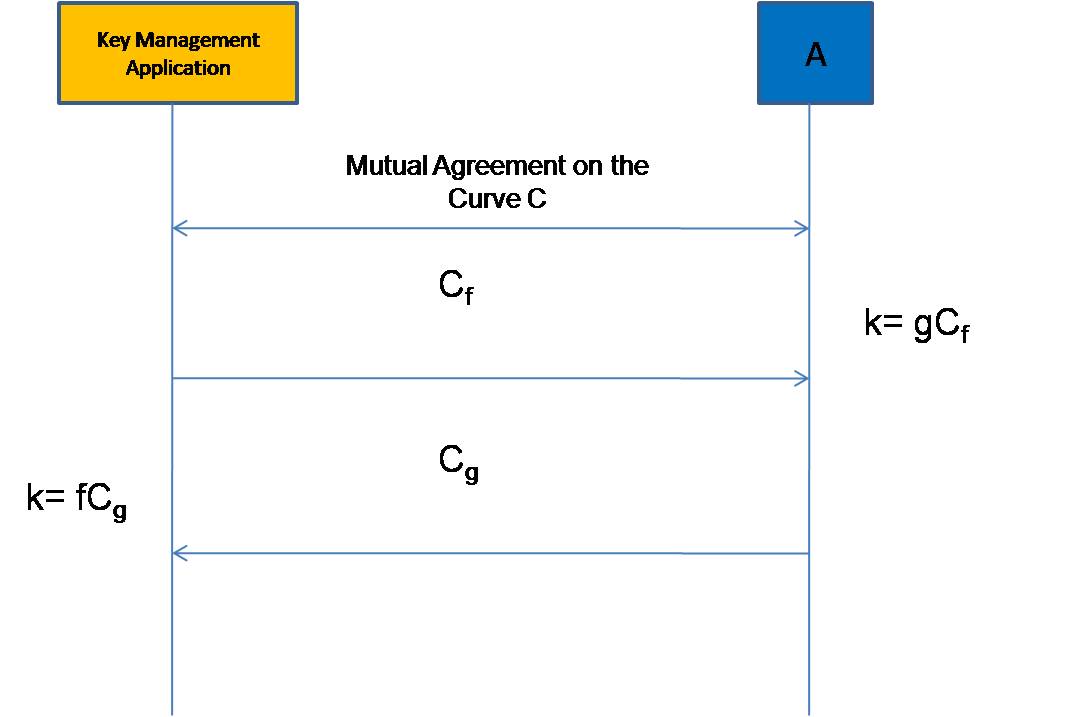}
	\caption{Key Distribution using ECDH}
	\label{fig:Figure12}
\end{figure}

Considering the IoT resource requirements, key management module will implement lightweight key distribution scheme. Recently, efforts in light weight key distribution schemes have received considerable attention from the research community \cite{Chung}, \cite{Lederer2009} ,\cite{Chung2008} and \cite{Akyildiz2010a}. Elliptic Curve Cryptography based Diffie Hellman (ECDH) is a hot candidate for a lightweight key distribution scheme in IoT. Apart from key distribution scheme the module will implement lightweight key generation algorithm and revocation. Lightweight revocation have received attention from the community such as in \cite{Mansour2014}. The Keys generated by the module will be used by the IoT nodes and security modules to implement security services in the IoT network. When a device join a network crypto keys for a device are generated and stored in storage. Upon revocation update takes place replacing the keys in the storage. Figure 11. shows the outline of key management module. Key management module process is as follows.

\textbf{Step 1:} Device joins the network and get registered.

\textbf{Step 2:} Key generation request is then initiated via SDN controller for a device from key management application.

\textbf{Step 3:} Crypto keys are generated and stored in the storage for a device.

\textbf{Step 4:} Generated keys are distributed to the devices using a key distribution algorithm.

\textbf{Step 5:} Key renewal and revocation process is carried out on the recommendations of flow analyzer to revoke list of nodes. Keys stored in the storage are expired and any data encrypted with the expired keys will not be considered valid.

\textbf{Step 6:} Keys generated are stored in the storage and distributed to the devices using a distribution algorithm.

Key generation process is as follows. \\

\textbf{Step 1:} Key management module and nodes for ECDH in the network agree on curve C. For each exchange between node (A) and node (B), node (A) generates a secret number f. This secret number f is used to compute corresponding public point which is computed as \[U=(q_f,w_f) = C_f\] B generate a secret number g with the use of which a corresponding public point is computed \[U=(q_g,w_g) = C_g\].

Key distribution scheme based on ECDH is as follows. Figure 12. shows how the process works.

\textbf{Step 1:} Compute the keys for nodes A and B, \[U=(q_f,w_f) = C_f\] and \[U=(q_g,w_g) = C_g\] by using elliptic curve cryptography.

\textbf{Step 2:} \[ C_f  and  C_g\] are exchanged over an insecure channel. Both the nodes can compute the shared secret as k = \[fC_g = gC_f\].

\subsection{IoT and Service Access Authentication}

\begin{figure}
	\centering
    \includegraphics[width=1.0\linewidth]{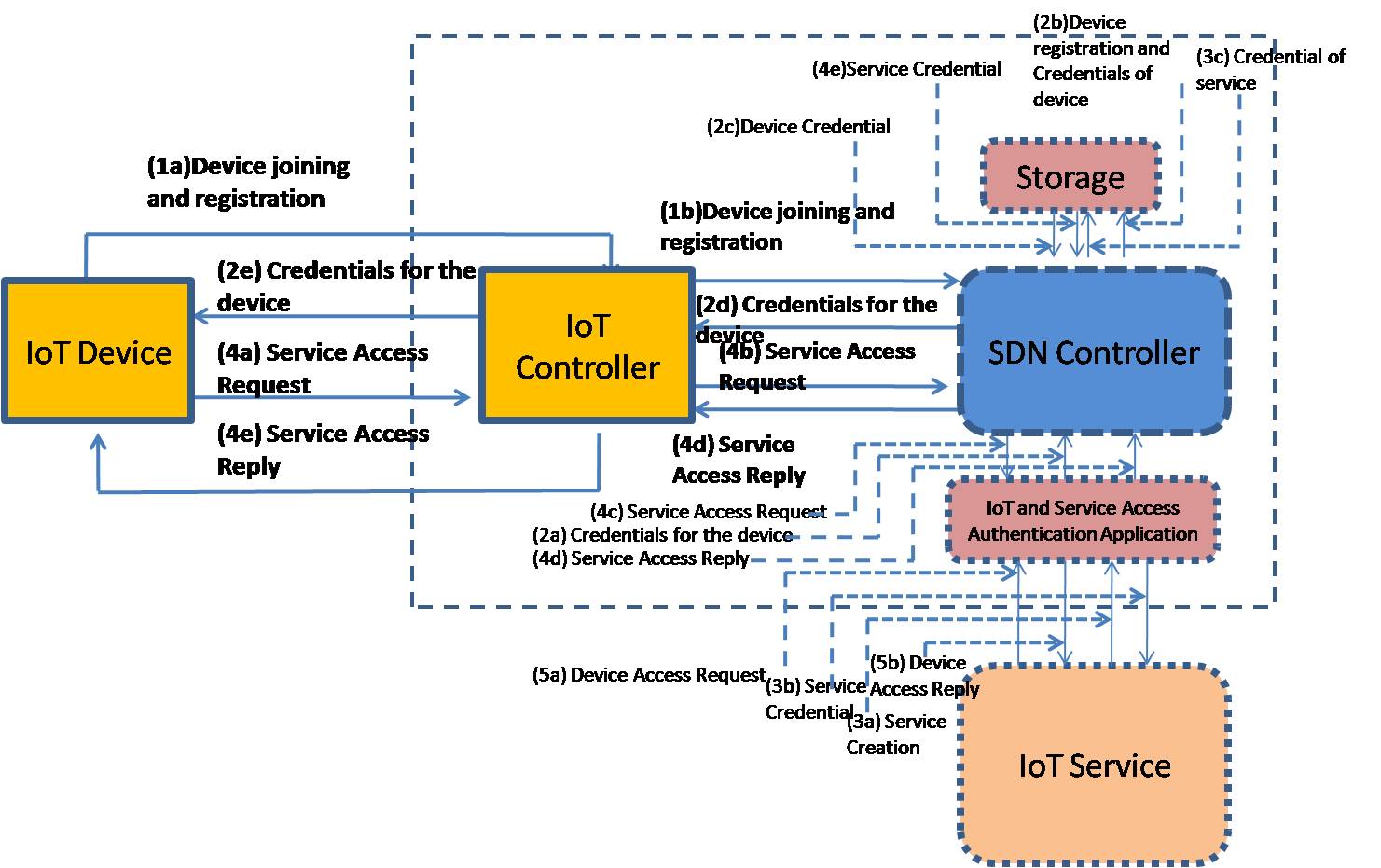}
	\caption{IoT and Service Access Authentication Module}
	\label{fig:Figure 13}
\end{figure}

\begin{figure}
	\centering
    \includegraphics[width=1.0\linewidth]{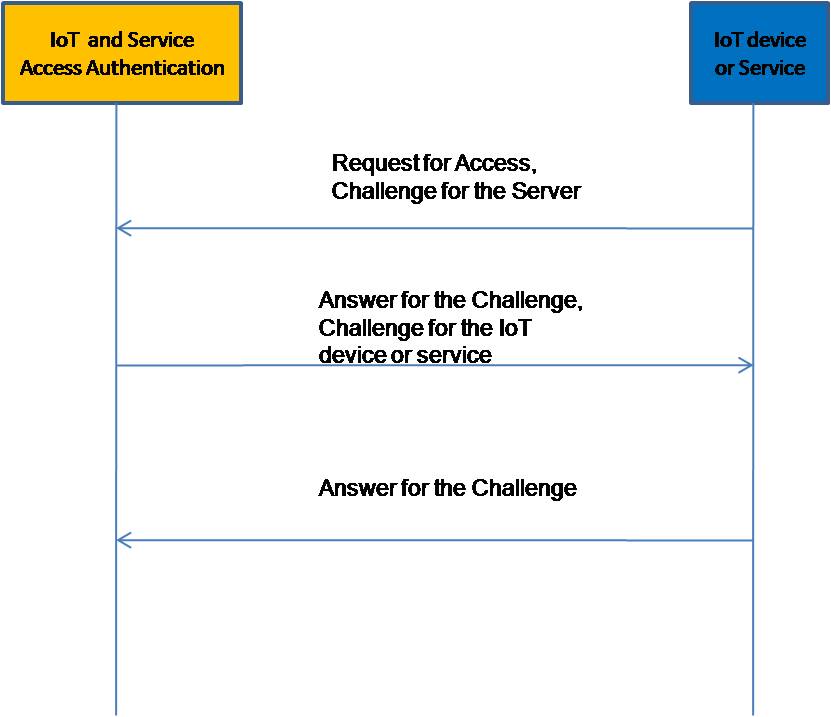}
	\caption{Authentication Process}
	\label{fig:Figure 14}
\end{figure}

IoT nodes and services will use the associated service crypto keys to access. The authentication algorithm will use the keys to authenticate the IoT nodes during the join time and any network element accessing the IoT service. There are various authentication algorithms for WSN proposed in the literature such as in \cite{Delgado-Mohatar2011}, \cite{ Rajeswari2016}. The keys to authenticate the access of the IoT services are called service access keys to ensure legitimate access of the IoT nodes by the service. A mutual authentication algorithm is used that authenticates the two parties at the same time using a challenge response authentication protocol. Figure 13. shows the outline of IoT and service access authentication module. The IoT and service access authentication process is as follows.

\textbf{Step 1:} Device upon joining get registered

\textbf{Step 2:} Crypto keys to access the service by the IoT nodes are distributed to the devices.

\textbf{Step 3:} IoT service wishing to access IoT nodes in future requests the IoT and service access authentication application to register and pass on the credentials to IoT services.

\textbf{Step 4:} When a service is requested by the IoT devices the authentication algorithm authenticates the devices.

\textbf{Step 5:} On the contrary if the device access is requested by the IoT services then the authentication algorithm authenticates the services based on the credential given to the service in step 3.

Authentication algorithm is shown in figure 14.

\textbf{Step 1:} When the network is deployed key distribution algorithm as described in the key management section will distribute the keys to the devices. Likewise, any IoT service wishing to gain access to IoT nodes will have the credentials by using key distribution algorithm.

\textbf{Step 2:} An IoT device or service wishing to gain access sends a challenge using credentials as given in step 1 to the IoT and service access authentication process. The process responds by answering the challenge using the credentials from step1 at the same time sending a challenge back to the IoT device.

\textbf{Step 3:} IoT device or service verifies the answer and at the same time responds to the challenge given by the IoT and service access authentication module using the credentials from step1.

\textbf{Step 4:} The IoT and service access authentication module verifies the answer from the IoT device or service, hence authenticating the device using the credentials from step1.

\subsection{Access Control}

\begin{figure}
	\centering
    \includegraphics[width=1.0\linewidth]{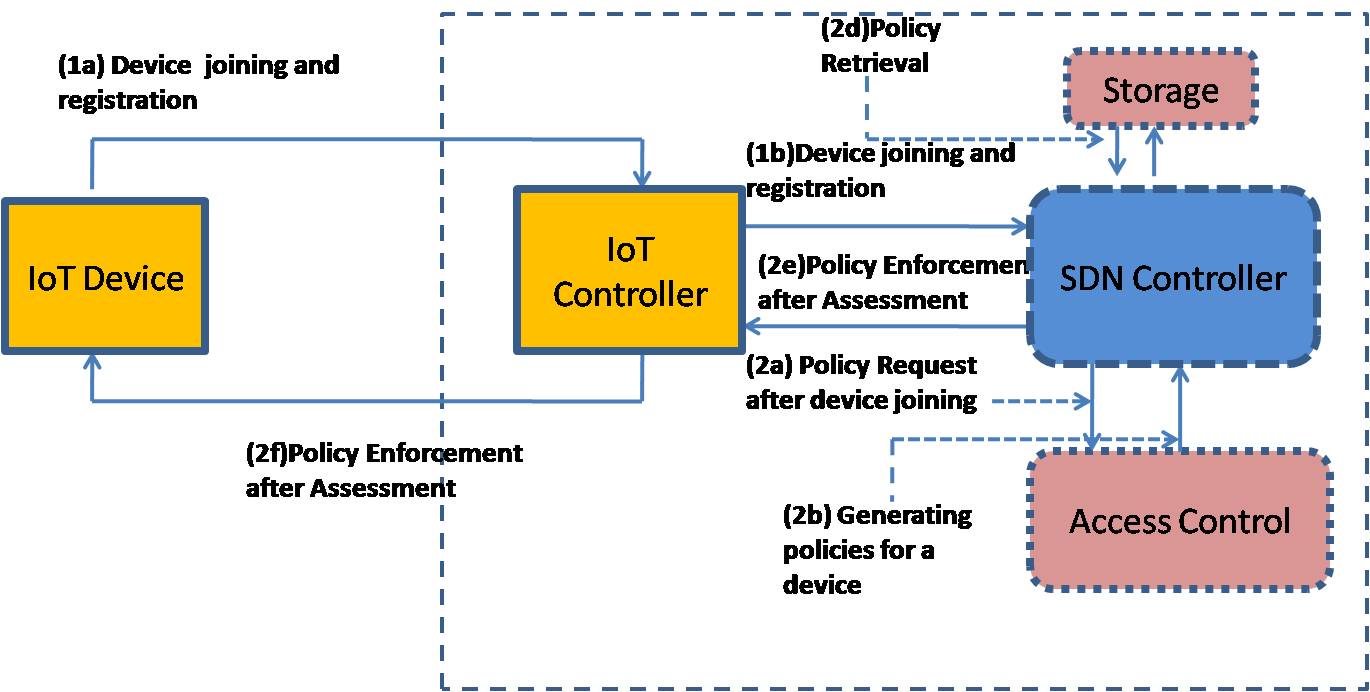}
	\caption{Access Control Module}
	\label{fig:Figure 15}
\end{figure}

\begin{figure}
	\centering
    \includegraphics[width=1.0\linewidth]{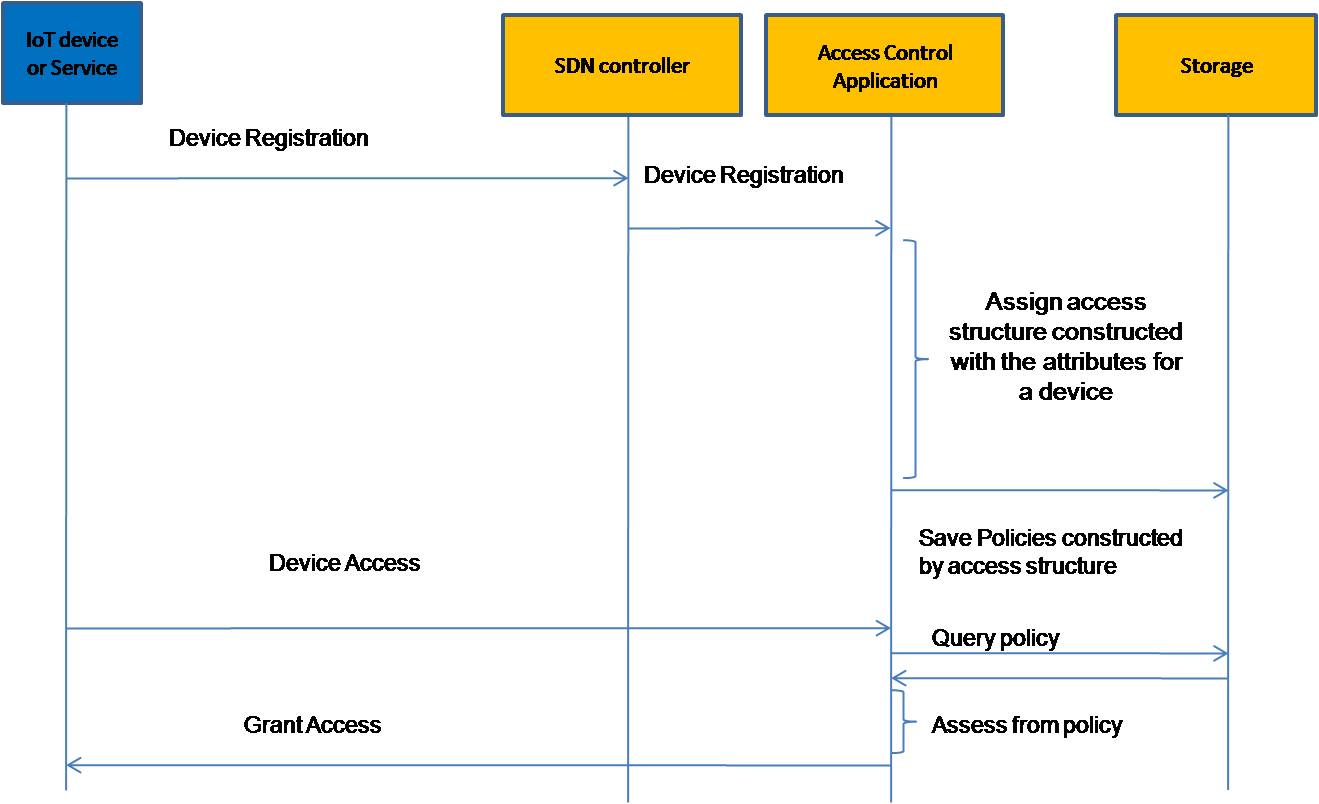}
	\caption{Access Control Process}
	\label{fig:Figure 16}
\end{figure}

The gateway will have a database of access control policies which are to be followed by the IoT nodes and the flows generated during the registration process. Access control algorithm will authorize, authenticate, approve access of the IoT resources based on access control policies. ABAC ( Attribute based access control ) is employed for IoT. ABAC is used in IoT for access control which can be seen in \cite{Lim2016}. During the registration process access structure based on attributes is assigned to the device joining the network. SDN controller assists the access control module to implement the access control by constructing attribute access structure on the flows generated by the devices. Figure 15. shows the outline of access control module. Access control module process is as follows.

\textbf{Step 1:} When a device joins and registers, the policies are constructed on the basis of access structure assigned by the access control module

\textbf{Step 2:} Policies for a device are generated and stored in the storage.

\textbf{Step 3:} Based on the policies generated for a device, policies are enforced on the devices by the SDN controller via IoT controller. These policies influences the IoT devices access operations on IoT services.

Access control algorithm is shown in Figure 16. The process is as follows.

\textbf{Step 1:} Upon initiation of device registration request the access control application initiates policy creation process.

\textbf{Step 2:} Access structure is assigned to a device which is constructed by using access tree derived from attributes for a particular flow or user in the network.

\textbf{Step 3:} Policy on the basis of access structure is derived and stored in the storage.

\textbf{Step 4:} Flow initiated from a device accesses IoT network which is then regulated by the policies stored in the storage. The flow is granted access influenced by the policy.

\subsection{Security Attack Mitigation Agent}

\begin{figure}
	\centering
    \includegraphics[width=1.0\linewidth]{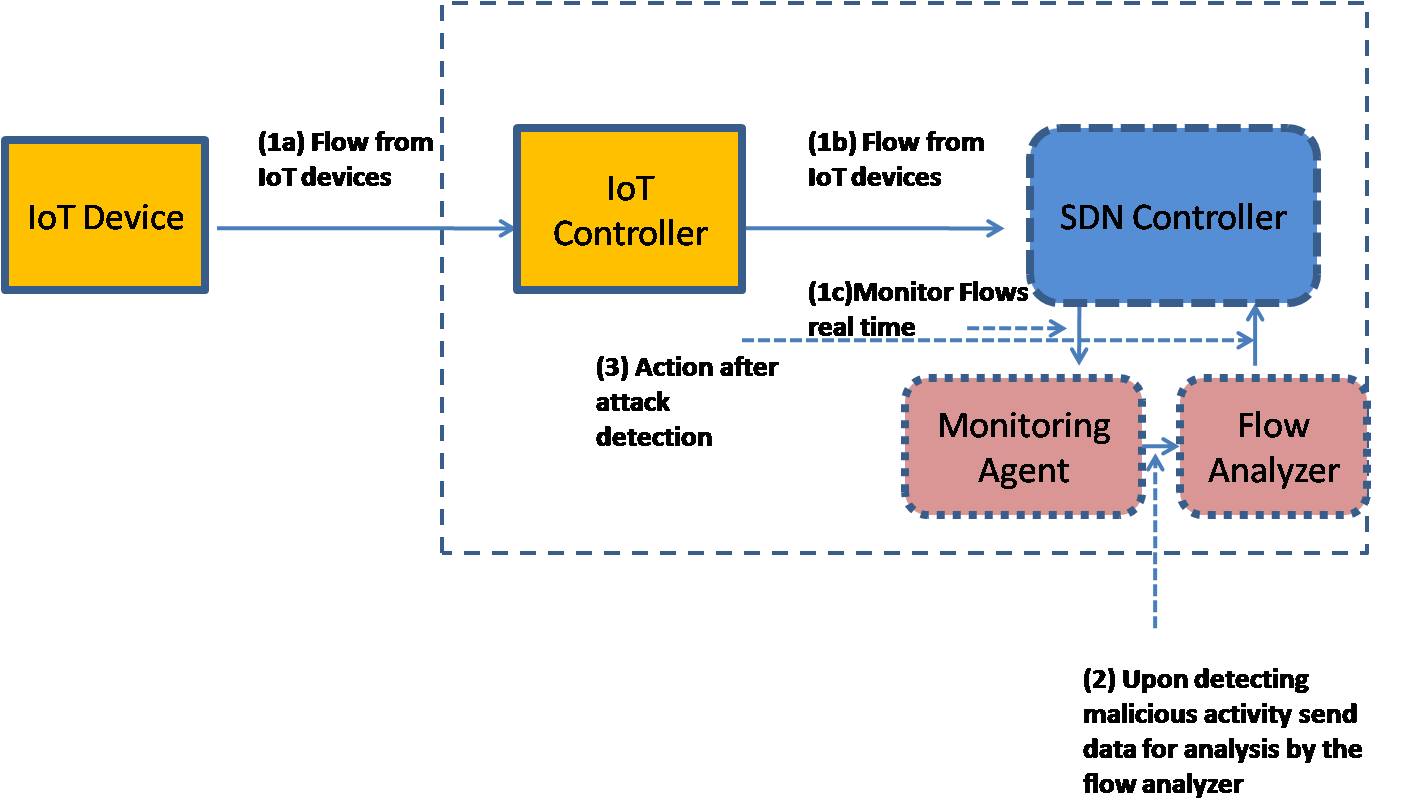}
	\caption{Security Attack Mitigation Module}
	\label{fig:Figure 17}
\end{figure}

\begin{figure}
	\centering
    \includegraphics[width=1.0\linewidth]{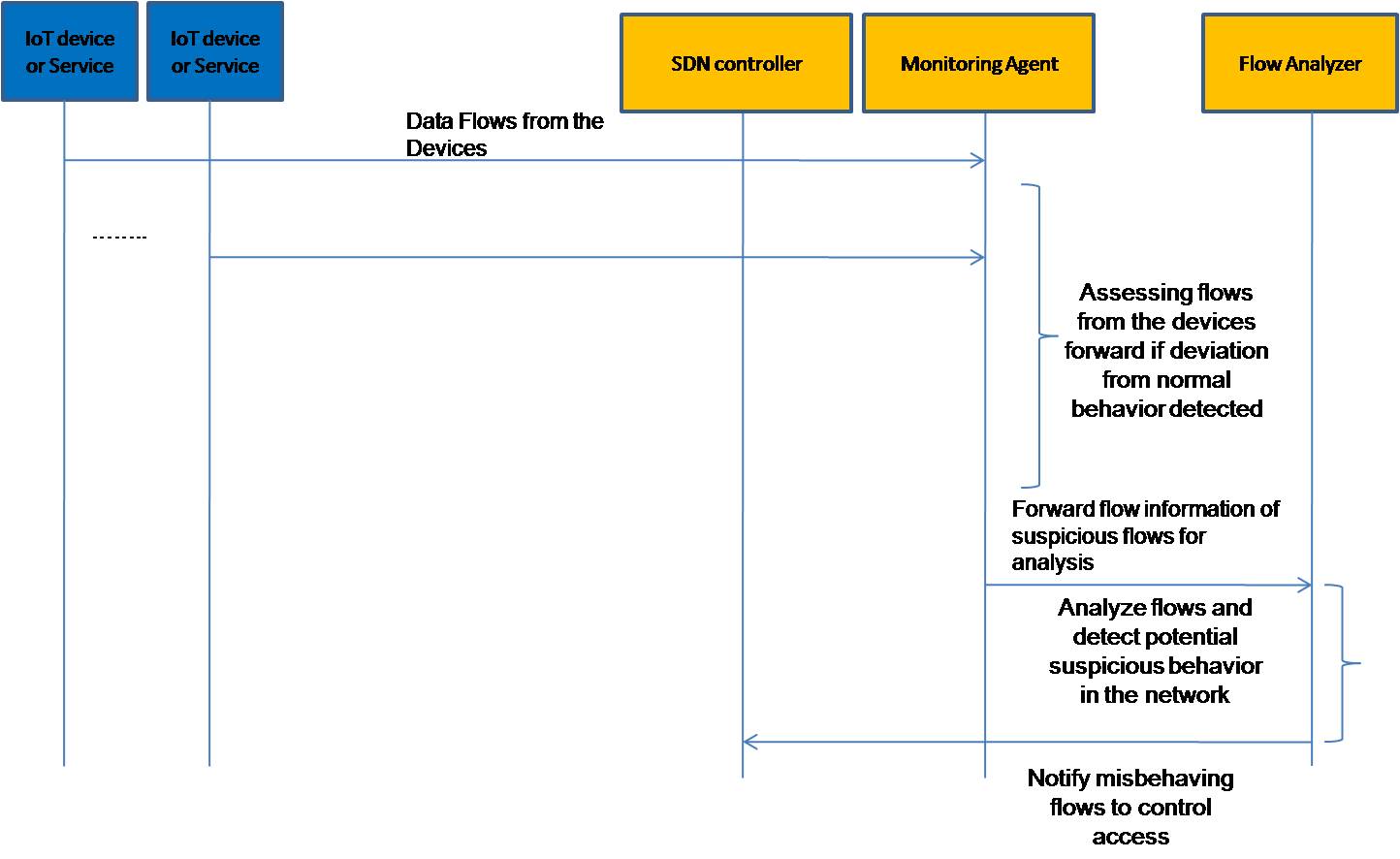}
	\caption{Security Attack Mitigation Process}
	\label{fig:Figure18}
\end{figure}

The algorithm in the mitigation agent will use the current status of the IoT network and its flows to detect possible malicious activity in the network. Threshold based mechanism will be used for detecting attacks on the IoT nodes and services. Type of attacks detected by the agent will be network scan, spoofing, injection, impersonation, DoS and DDoS attacks. Besides a database of access control policies, there is a database for auditing the flows in the IoT network. This is maintained by monitoring agent with the assistance of SDN controller. The audit trails of the IoT flows will be used by the flow analyzer to detect malicious intrusion of the attackers in the network. It will also propose actions to be taken by the SDN controller with regards to the flows from IoT devices. Figure 17. shows the outline of security attack mitigation agent module. Security attack mitigation agent module works as follows.

Step 1: Flows from the IoT devices via IoT controller are accounted by the SDN controller which are monitored by the monitoring agent.

Step 2: Auditing of the flows is taken care of by the flow analyzer which proposes actions or defenses against a possible malicious operation by the devices. SDN controller then implements strategies to counter malicious activities by the flows in future.

Attack mitigation process is shown in Figure 18. The process is as follows

\textbf{Step 1:} Data flows from the devices are monitored in real time by the monitoring agent.

\textbf{Step 2:} Upon detection of deviation from normal behavior of any of the flow, Flow analyzer is notified about the deviated flows by the monitoring agent.

\textbf{Step 3:} After analysis by the flow analyzer, the SDN controller is notified to limit access of the malicious devices.
\begin{table*}[htbp]
	\centering
	\caption{Summary of Security Services Provisioned by Proposed Management Framework}
	\scalebox{0.85}{
	\begin{tabular}{lrrrlrr}
		\toprule
		& \multicolumn{1}{l}{Privacy Module} & \multicolumn{1}{l}{Trust Module} & \multicolumn{1}{l}{Key Management Module} & IoT and Service Access Authentication & \multicolumn{1}{l}{Access Control } & \multicolumn{1}{l}{Security Attack Mitigation Agent} \\
		\textbf{Security Attacks, Issues and Threats in IoT} &       &       &       &       &       &  \\
		\midrule
		Confidentiality and Privacy & \multicolumn{1}{l}{               \checkmark} &       & \multicolumn{1}{l}{                      \checkmark} &       &       &  \\
		\midrule
		Handling Security Attacks &       &       &       &       &       & \multicolumn{1}{l}{                               \checkmark} \\
		\midrule
		Authentication in IoT &       &       & \multicolumn{1}{l}{                      \checkmark} &                                 \checkmark &       &  \\
		\midrule
		Identity Spoofing Attack &       &       &       &       &       & \multicolumn{1}{l}{                               \checkmark} \\
		\midrule
		Access Control in IoT &       &       &       &                                  & \multicolumn{1}{l}{     \checkmark} &  \\
		\midrule
		Trust in IoT &       & \multicolumn{1}{l}{                \checkmark} &       &       &       &  \\
		\midrule
		Attacks on Availability &       &       &       &       &       &  \\
		\midrule
		Impersonation Attacks &       &       &       &       &       & \multicolumn{1}{l}{                              \checkmark} \\
		\midrule
		Eavesdropping & \multicolumn{1}{l}{      \checkmark} &       & \multicolumn{1}{l}{        \checkmark} &       &       & \multicolumn{1}{l}{                              \checkmark} \\
		\midrule
		Data Corruption & \multicolumn{1}{l}{      \checkmark} &       & \multicolumn{1}{l}{        \checkmark} &       &       &  \\
		\midrule
		Data Modification & \multicolumn{1}{l}{      \checkmark} &       & \multicolumn{1}{l}{        \checkmark} &       &       &  \\
		\midrule
		Secure Routing and Forwarding in IoT &       &       &       &            \checkmark &       & \multicolumn{1}{l}{           \checkmark} \\
		\midrule
		Robustness and resilience management in IoT &       &       &       &       &       & \multicolumn{1}{l}{           \checkmark} \\
		\midrule
		Audit Control for IoT &       &       &       &       &       & \multicolumn{1}{l}{           \checkmark} \\
		\midrule
		Secure Network Access &       &       & \multicolumn{1}{l}{        \checkmark} &            \checkmark &       &  \\
		\midrule
		Secure Storage &       &       &       &       & \multicolumn{1}{l}{       \checkmark} &  \\
		\midrule
		Tamper Resistance & \multicolumn{1}{l}{     } &       & \multicolumn{1}{l}{        } &       & \multicolumn{1}{l}{                   \checkmark} &  \\
		\midrule
		User Identification and Identity Management &       &       &       &            \checkmark &       &  \\
		\bottomrule
	\end{tabular}%
	\label{tab:addlabel}%
}
\end{table*}%

\section{Summary of IoT security requirements handling by proposed management framework}
\label{sec:solutions}

This section describes how the proposed management framework handles IoT security issues, requirements, threats and attacks. Table 1. summarizes what security services are fulfilled by the proposed framework. The modules are listed in the column of the table while security attacks, issues and threats in IoT are listed in the rows of the table. The table helps in assessing which modules in the proposed framework fulfills what security requirements, issues and threats of IoT in the network discussed earlier. Discussion on solutions and recommendations are as follows.

\subsection{Confidentiality and Privacy}

The SDN based security controller is responsible to facilitate confidentiality and privacy of data exchanged among the IoT devices. Privacy module together with key management module in the security framework is responsible for ensuring protection against eavesdropping, data corruption and modification across the IoT network. Key management module is responsible for managing cryptographic materials used by the privacy module.

\subsection{Handling Security Attacks}

The Security attack mitigation agent is responsible for regulating traffic within the IoT network. Constant monitoring of the traffic flows along with the analyzer helps in detecting DoS, DDoS, injection, spoofing and impersonation attacks.

\subsection{Authentication in IoT}

Authentication in IoT is ensured by IoT and service access authentication together with key management module in the security framework. The module ensures that the joining device accessing service from the IoT network is authenticated and authorized. Key management module is responsible for managing cryptographic materials used by the IoT and service access authentication module.

\subsection{Access Control in IoT}

Access control module in IoT implements security policies in the IoT network. Hence, the module ensures that IoT device perform authorized action while it is a part of the network. Illegtimate access should be denied by the access control module.

\subsection{Trust in IoT}

Trust module ensures trusted communication between IoT devices and IoT network. An IoT device or service reputation is downgraded if negative experiences are encountered over time. Hence, historical actions are taken into account to establish trusted linkage between the IoT devices and network.

\subsection{Eavesdropping}
Privacy module together with key management module in the proposed framework ensures that the communication between IoT devices and the gateway takes place in a secure manner. This ensures that no malicious entity eavesdrop on the ongoing communication. Key management module manages cryptographic keys used for ensuring protection against eavesdropping.

\subsection{Data Corruption}
Corruption of data is prevented by the privacy module which work with key management module. The module implements cryptographic algorithm to ensure that no malicious entity can corrupt the data generated from the IoT devices. Cryptographic keys used are managed by key management module.

\subsection{Data Modification}
Illegal modification of the data is ensured by privacy module which works with key management module. Cryptographic algorithm enables secure communication which disallows any malicious nodes to modify the data originated from IoT devices. Key management module is responsible for managing cryptographic keys used by privacy module to ensure protection against illegal modification.

\subsection{Secure Routing and Forwarding in IoT}
IoT and service access authentication module and security attack mitigation agent ensures that the data is routed and forwarded in the IoT network in a secure manner. Any routing attacks are detected by the security attack mitigation agent hence ensuring secure data forwarding.
IoT and service access authentication module filters legitimate nodes to become of the IoT network.

\subsection{Robustness and Resilient Management in IoT}
Security attack mitigation agent detects any kind of malicious attacks which halts the services offered by IoT network. Also, the agents takes appropriate actions to alleviate the situation hence ensuring that the IoT network runs in a normal manner.

\subsection{Audit Control in IoT}
Security attack mitigation agent monitors the flows in the IoT network which helps in auditing actions. If any inappropriate action is observed the agents reacts to the situation and ensures smooth operation of the network.

\subsection{Secure Network Access}
IoT and service access authentication module together with key management module authenticates all the joining nodes hence ensuring only validated ones become part of the network. Also, access control ensures only legitimate actions are carried out by the nodes in the IoT network. Key management module is responsible for managing cryptographic keys used by IoT and service access authentication module.

\subsection{Secure Storage}
Access control module ensures secure access of the storage as only legitimate nodes are able to access the data. Furthermore, these legitimate nodes perform only legal actions on the data stored in the node.

\subsection{Tamper Resistance}
Access control module ensures any illegal tampering of the IoT nodes in the network.

\subsection{User identification and identity management}
IoT and service access authentication module is responsible for validating the joining nodes hence verifying the identity of the nodes.

\section{Conclusion}

Advancement in programmable networks have enabled novel paradigm of SDN which have opened opportunities of easing management of networks. Emerging interconnected embedded devices paradigm of IoT is different than conventional wired networks which are usually constrained in resources. Hence, managing such type of network raises challenges which are of unique nature. In this paper, management challenges of IoT are identified and discussed . One of the aspect of management solution for IoT is security provisioning. In this paper, security management of IoT is dealth with. Security threats, attacks, issues and requirements in IoT are discussed which need attention from the researchers. Lately, potentials of SDN to manage IoT is been investigated. It is no doubt that SDN paradigm offer an excellent opportunity to assure security in IoT as security control will be centralized.

Hence, in this paper, management framework based on SDN principles for provisioning security services in IoT is proposed. Proposed security controller consists of privacy, trust, key management, IoT and service access authentication and security attack mitigation agent module. Privacy module ensures privacy is preserved in IoT. Trust module makes sure that the communication in IoT network takes place in a trusted environment. Key management module handles key generation and revocation in IoT network. IoT and service access authentication authenticates nodes and services within IoT network. Security attack mitigation agent detects attacks in the network and takes countermeasure actions to prevent attacks. In the end, how the security attacks and threats in IoT are handled by the proposed SDN based framework is discussed. In the future, each module will be implemented and evaluated in the framework with respect to overall overheads and resource consumption.


%



\ifCLASSOPTIONcaptionsoff
\newpage
\fi



\bibliographystyle{IEEEtran}
\bibliography{IoTproposal}
%



%





\end{document}